\documentclass[toc]{PoS}
\usepackage{graphicx}
\usepackage{epsfig}
\newcommand\be{\begin{equation}}
\newcommand\ee{\end{equation}}
\newcommand\ignore[1]{}

\title{Hybrid Monte Carlo simulation  on the graphene hexagonal  lattice}

\ShortTitle{Monte Carlo simulation of graphene}

\author{Richard Brower\footnote{Speaker}\\
ECE and Physics Departments and Center for Computational Science\\
Boston University, Boston, MA 02215\\
E-mail: \email{brower@bu.edu}}

\author{Claudio Rebbi$^*$ \\
Physics Department and Center for Computational Science\\
Boston University, Boston, MA 02215\\
E-mail: \email{rebbi@bu.edu}}

\author{David Schaich \\
Physics Department\\
University of Colorado, Boulder, CO 80309\\
E-mail: \email{David.Schaich@Colorado.EDU}}

\abstract{One of the many remarkable properties of graphene is that in
  the low energy limit the dynamics of its electrons can be
  effectively described by the massless Dirac equation. This has
  prompted investigations of graphene based on the lattice simulation
  of a system of 2-dimensional fermions on a square staggered lattice.
  We demonstrate here how to construct the path integral for graphene
  working directly on the graphene hexagonal lattice. For the
  nearest neighbor tight binding model interacting with a long range
  Coulomb interaction between the electrons, this leads to the hybrid
  Monte Carlo algorithm with no sign problem. The only approximation
  is the discretization of the Euclidean time. So as we extrapolate to
  the time continuum limit, the exact tight binding solution maybe found
  numerically to arbitrary precession on a finite hexagonal lattice.
  The potential for this approach is tested on a single hexagonal
  cell. }

\FullConference{The XXIX International Symposium on Lattice 
Field Theory - Lattice 2011
July 10-16, 2011
Squaw Valley, Lake Tahoe, California}
\begin{document}
\section{Introduction}
\label{sec:intro}

During the last few years  graphene, a single layer of carbon atoms forming 
a hexagonal lattice, has burst on the scientific scene as a remarkable 
material with intriguing theoretical properties and potentially
astounding technological applications~\cite{graphene}.  Its electronic 
structure is well approximated by the tight binding Hamiltonian $H$ which describes
the quantum mechanical motion of the electrons, one per atom at 
half-filling, that are not part of the cloud of electrons responsible
for the covalent binding of the carbon atoms.   The simplest model Hamiltonian, 
$H = H_2 + H_C$,  consists of two terms, a quadratic Hamiltonian $H_2$ which describes 
the hopping of the electrons between neighboring atoms, and a
Hamiltonian $H_C$ which describes the Coulomb repulsion of the
electrons.  In the the limit, where the strength
of the hopping term is taken to be much larger than strength of
the Coulomb interaction, the quadratic Hamiltonian $H_2$
gives origin to a dispersion formula which for
low momenta is analogous to the dispersion formula for relativistic
fermions in two dimensions~\cite{dispersion}.  This has prompted some 
researchers to adapt to graphene lattice gauge theory techniques 
which have been profitably used for the study of Quantum Chromodynamics 
and other particle systems~\cite{LGT}.  In the work of~\cite{LGT-graphene, Lahde:2011yu}
one approximates the
tight-binding Hamiltonian of graphene with a Dirac Hamiltonian,
incorporates the Coulomb interaction through the introduction of a suitable
electromagnetic field, and finally discretized the resulting continuum
quantum field theory on a hypercubic space-time lattice.  The hybrid
Monte Carlo method~\cite{HMC}, widely used in lattice 
gauge theory to simulate fermions interacting
with quantum gauge fields, can then be used to investigate
the graphene system.

The approach outlined above has led to very interesting and valuable
results~\cite{LGT-graphene,Lahde:2011yu}, yet  it has several limitations. First the use
of staggered fermion on a cubic lattice looses the direct connection to 
special symmetries of hexagonal lattice and introduces
symmetry breaking artifacts, know in lattice QCD as taste
splitting~\cite{Giedt:2009td}. Second a 4-d lattice is required to
simulate the coulomb potential which is not only computational expensive but
distorted at the scale of the physical spatial  lattice spacing. One would think that, since the
starting point is a system already defined on a lattice, it should be
possible to apply the hybrid Monte Carlo technique directly to the
graphene lattice. The clear advantage of working on the hexagonal graphene  lattice is the
direct connection to the experimentally determined physical lattice
constants of the tight-binding model, which represents
an accurate description of the experimental system. Here we wish to
illustrate how this can be done.

\section{The graphene lattice}
\label{sec:lattice}

\begin{figure}[ht]
\begin{minipage}[b]{0.47\linewidth}
\centering
\includegraphics[width=7.3cm]{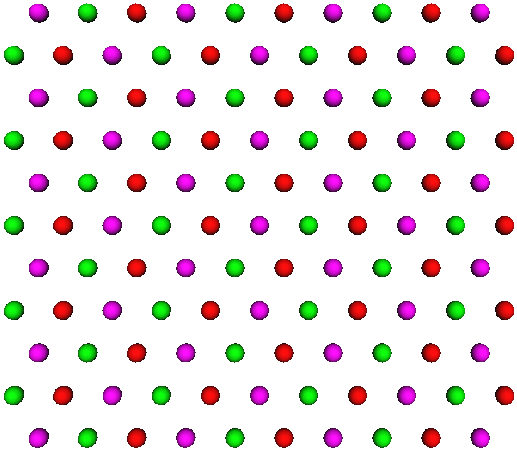}
\caption{The underlying triangular lattice.}
\label{fig:figure1}
\end{minipage}
%\hspace{0.5cm}
\hspace{0.06\linewidth}
\begin{minipage}[b]{0.47\linewidth}
\centering
\includegraphics[width=7cm]{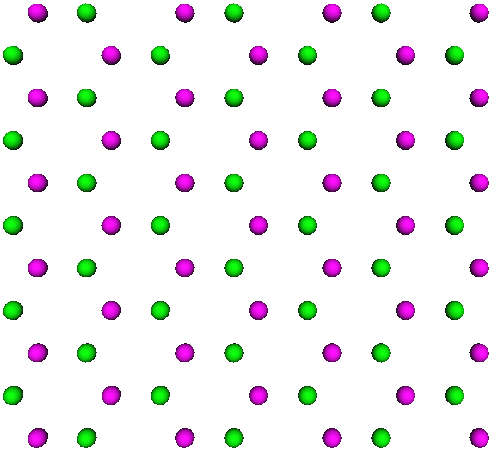}
\caption{The hexagonal graphene lattice.}
\label{fig:figure2}
\end{minipage}
\end{figure}

It is useful to review the geometrical properties of graphene.
Graphene is a system of interacting electrons located at the vertices
of a hexagonal lattice. It is convenient to think of the graphene
lattice as consisting of two triangular sublattices, which we denote
by $A$ and $B$, which together with the centers of the hexagons
(sublattice $C$) form a finer, underlying triangular lattice. In
Figures~\ref{fig:figure1} and ~\ref{fig:figure2}, we illustrate the
full underlying triangular lattice and the hexagonal lattice which one
obtains by eliminating the sites of one of the three sublattices. $A$
and $B$ are Bravais lattices so given any two primitive vectors $\vec
a_1, \vec a_2$ their sites ( at $\vec r_A = x_1 \vec a_1 + x_2 \vec
a_2$ and $\vec r_B = x'_1 \vec a_1 + x'_2 \vec a_2$ respectively) are
conveniently enumerated by two integers $(x_1,x_2)$.

\begin{figure}[t!]
%\vspace{2.5cm}
\begin{minipage}[b]{0.47\linewidth}
\centering
\includegraphics[width=0.8\textwidth]{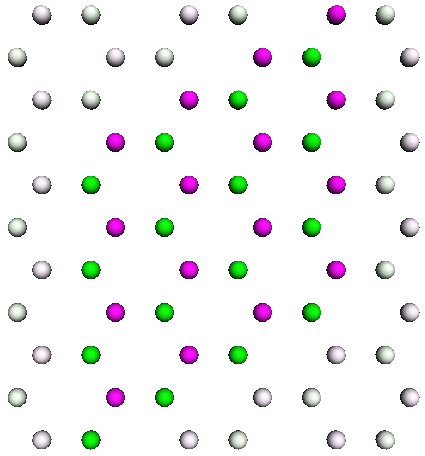}
\caption{Periodic lattice with $2 L_1 L_2$ sites viewed as parallelogram subdomain
in the plane illustrated for period $L_i = 4$  along two primitive vectors:
$\hat a_1 \sim \sqrt{3} \hat x/2 +  \hat y/2, \; \hat a_2 \sim \hat y$,
defined relative to the Cartesian unit vectors, $(\hat x,\hat y)$.}
\label{fig:figure3}
\end{minipage}
\hspace{0.2cm}
\raisebox{0.5cm}{
\begin{minipage}[b]{0.47\linewidth}
\centering
\includegraphics[width=0.9\textwidth]{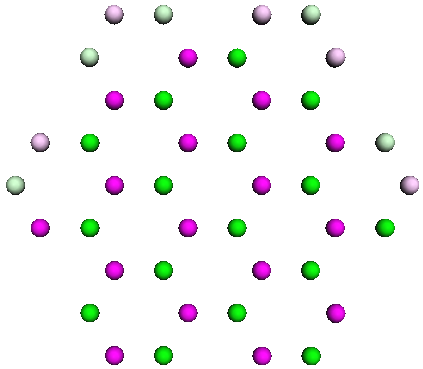}
\caption{The sublattice  with period $L=4$ from Figure~\protect\ref{fig:figure3} is
mapped into periodic hexagonal domain by moving four vertices to equivalent sites
combined with ten image or ghost sites shown in lighter color to complete the pattern.}
\label{fig:figure4}
\end{minipage}}
\end{figure}

One may impose periodic boundary conditions (PBC) by identifying sites
that differ by a translation by $L_i$ along the primitive vectors:
$x_i \simeq x_i + L_i$. In Figure~\ref{fig:figure3},  we show an example of PBC for $L_1 = L_2 =4$
with  green and purple color  for the independent lattice sites on each sublattice.  In this
representation, each sublattice A and B is  mapped into a  rectangle with periodic boundary 
condition very easily coded by two dimensional arrays.  The pattern of these sites does not
look reminiscent of the hexagonal structure of graphene.   However, it
is equivalent. If one rearranges the sites by replacing some of the vertices with
their periodic images and one adds to these vertices a few of their periodic
images  one obtains a pattern of sites
which clearly shows the hexagonal structure of graphene with
3 fold rotational symmetry as illustrated in Figure~\ref{fig:figure4}. Of course, 
one could  also introduce different periods $L_1, L_2$ along two directions 
of the fundamental cell and consider other types of boundary conditions.
For example, by taking $L_2 \gg L_1$ with PBC along the shortest 
dimension and open boundary conditions direction along the other
one can simulate a nanotube.  
 
\section{The dynamics}
\label{sec:dynamics}

We introduce fermionic annihilation and
creation operators $a_{x,s}, a^\dag_{x,s}$ for the electrons on the
two sublattices, where $x$ is a site index for both the A and B sublattices and $s= \pm 1$
labels the spin of the electrons.
The Hamiltonian $H = H_2 + H_C $ consists of two terms.  The quadratic
kinetic term is
\begin{equation}
H_2=\sum_{\langle x,y\rangle,s}  - s\kappa (a^\dag_{x,s}a_{y,s} + \mbox{h.c.}),
\label{eq2}
\end{equation}
where the sum runs over all pairs $\langle x,y\rangle$ of
nearest neighbor sites (coupling the $A$ and $B$ sublattices) and 
the two values of the spin.
The Coulomb interaction term is
\begin{equation}
H_C=\sum_{x,y} e^2 V_{x,y} q_x q_y = \frac{1}{2} \sum_{x \ne y}  q_x \frac{e^2}{\sqrt{(\vec r_x - \vec r_y)}}  q_y  +  \sum_xq_x \frac{e^2}{r_0}  q_x
\label{eq3}
\end{equation}
where $q_x=a^\dag_{x,1} a_{x,1}+a^\dag_{x,-1} a_{x,-1} -1$ is a local
charge operator and $V$ is the interaction potential with physical
co-ordinates $\vec r_x, \vec r_y$ for each Carbon atom. Note that we
have introduced a -$1$ in the charge operator $q_x$ to account for the
background charge of the carbon ion: this ensures that the system is
neutral at half filling, and it will play an important role for our
functional integral formulation. Insofar as the matrix $V$ in
Eq.~\ref{eq3} is concerned, it could be the actual 3d Coulomb
potential, but could be any other interaction potential. The only
constraint is that the matrix $V_{x,y}$ must be positive definite.

The phenomenological constants are a nearest neighbor lattice spacing
of $a \simeq 1.42\AA$, $\kappa \simeq 2.8 Mev$ and $e^2$ is effective
parameter for $e^2/\hbar v \simeq 300 \times e^2/\hbar c$ so that
unlike free electrons effective charge is very strong and the
electrodynamics essential static, free of magnetic effects. The single
site Coulomb~\cite{Wehling:2011} radius, $r_0 \simeq 0.5 a $, is
essential for stability of the vacuum
configuration. We also note that in Eq.~\ref{eq2}
we have neglected the smaller next-to-nearest neighbor hopping within
each sublattice , estimated  to be $ \kappa' \simeq 0.03 \kappa$. This would
introduce a (probably manageable) complex phase in the path integral.

Finally, we observe that the Hamiltonian of Eqs.~\ref{eq2}-\ref{eq3} 
commutes with the angular momentum generators,
\begin{eqnarray}
  J_{\pm} & = & a^\dag_{x,s}\sigma^{ss'}_{\pm} a_{x,s'}, \nonumber \\
  J_3 & = & a^\dag_{x,s} \sigma^{ss'}_3 a_{x,s'}/2 \; , 
  \label{eq:iso}
\end{eqnarray}
which act on the spin of the electron and play the
role in the tight-binding model of an internal or  ``isospin'' symmetry.
Additional symmetries of the system are the overall momentum conservation,
parity under reflection with respect to the crystal axes and under
interchange of the $A$ and $B$ sublattices, and 
the separate conservations of the number of electrons with spin up and 
spin down.

In order to explore the properties of the system one would like to calculate
expectation values
\begin{equation}
\langle {\cal O}_1(t_1) {\cal O}_2(t_2) \dots \rangle = Z^{-1} \mbox{Tr}\,
T[{\cal O}_1(t_1) {\cal O}_2(t_2) e^{\textstyle-\beta H}]  \; ,
\label{eq5}
\end{equation}
where $\cal O$ stands for a generic observable,
$\beta$ can be interpreted as an extent in Euclidean time,
$T[\dots]$ stands for time ordering of the operators inside the square
bracket with respect to the Euclidean evolution implemented by 
$\exp(-\beta H)$, and $Z=\mbox{Tr}\, \exp(-\beta H)$ is the partition
function.

\section{Path integral formulation}
\label{sec:pathIntegral}

Our goal is to provide an equivalent path integral formulation of
Eq.~\ref{eq5} conducive to calculation by numerical simulation,
following a rather standard procedure to convert from the Hamiltonian
into a Lagrangian.  We will first express the expectation values and
the partition function in terms of an integral over anticommuting
fermionic fields, i.e.~elements of a Grassmann algebra.  (See for 
example~\cite{Negele-Orland}.)  It is important that the 
Hamiltonian be normal ordered.  This is true of $H_2$ and of
the non-local terms in $H_C$.  However the normal ordered
$:H_C:$ differs from $H_C$ by a diagonal quadratic term which
can be added to $H_2$.  With this in mind one can proceed
to the path integral formulation, which gives origin
to an integral over the  Grassmann variables of an
exponential containing a quadratic form in the fermionic fields, from
$H_2$ and the normal ordering of $H_C$, as well as a quartic
expression from $:H_C:$.  The quartic expression can be reduced to a
quadratic form by a Hubbard-Stratonovich transformation~\cite{HST},
through the introduction of a suitable auxiliary bosonic field (in our
case a real field), and now the Gaussian integral over the fermionic
variables can be explicitly performed, leaving an integral over the
bosonic field only~\cite{BSS}.  The problem, however, is to obtain an
integral that can be interpreted as an integration over a well defined
probabilistic measure, which can be approximated by stochastic
simulation techniques.  In the following we show how this can
be accomplished by taking advantage of the symmetries of the system

We start by rewriting the expression for the charge as
\begin{equation}
q_x=a^\dag_{x,1} a_{x,1}-a_{x,-1} a^\dag_{x,-1}.
\label{eq6}
\end{equation}
We now introduce hole creation and annihilation operators for the
electrons with spin $-1$:
\begin{equation}
b^\dag_x = a_{x, -1}, \quad b_x=a^\dag_{x,-1}
\label{eq7}
\end{equation}
so that the charge becomes
\begin{equation}
q_x=a^\dag_x a_x-b^\dag_x b_x.
\label{eq8}
\end{equation}
Note that we dropped the spin indices since from now on $a, a^\dag$ and
$b, b^\dag$ will always refer to spin 1 and $-1$, respectively.
Finally we change the sign of the $b, b^\dag$ operators on one of the
sublattices.  The crucial constraint is that all redefinitions of the
operators respect the anticommutator algebra.
Because $H_2$ only couples sites on the
two different sublattices, it takes the form
\begin{equation}
H_2=   \sum_{\langle x,y\rangle} -\kappa (a^\dag_x a_y + b^\dag_x b_y 
+ \mbox{h.c.}).
\label{eq9}
\end{equation}

We introduce fermionic coherent states
\begin{equation}
\langle \psi^*, \eta^* \vert = \langle 0 \vert
e^{\textstyle-\sum_x (a_x  \psi_x + b_x \eta_x)}
\qquad
\vert \psi, \eta \rangle = 
e^{\textstyle-\sum_x (\psi_x a^\dag_x+\eta_x b^\dag_x)} \vert 0 \rangle, 
\label{eq10}
\end{equation}
where $\psi_x, \psi_x^*, \eta_x, \eta_x^*$ are anticommuting
fermionic variables (elements of a Grassmann algebra). The path integral formulation is obtained by factoring
\begin{equation}
e^{\textstyle-\beta H} =  e^{\textstyle- H\, \delta} e^{\textstyle- H\, \delta} \dots e^{\textstyle- H\, \delta}
\quad (N_t\; \mbox{terms})
\label{eq11}
\end{equation}
with $\delta = \beta /N_t$, and then inserting repeatedly among
the factors the resolution of the identity expressed in terms 
of an integral over the fermionic variables. The trace
in Eq.~\ref{eq5} must also be expressed in terms of a similar
integral (see e.g.~\cite{Negele-Orland} for details).  This leads to
integrals over fermionic fields $\psi_{x,t}, \psi_{x,t}^*, 
\eta_{x,t}, \eta_{x,t}^*$ (the index $t=0,\cdots,N_t-1$ appears because of the
multiple resolutions of the identity and can be thought of as an
index labeling Euclidean time), which contain in the integrand
expressions of the type
\begin{equation}
\langle \psi^*_{x,t}, \eta^*_{x,t} \vert e^{\textstyle-H \,\delta} \vert
\psi_{x,t}, \eta_{x,t} \rangle.
\label{eq12}
\end{equation}
The last ingredient is the identity
\begin{equation}
\langle \psi^*_{x,t}, \eta^*_{x,t} \vert F(a_x^\dag,b_x^\dag,a_x,b_x) 
\vert \psi_{x,t}, \eta_{x,t} \rangle = 
F(\psi_{x,t}^*, \eta_{x,t}^*, \psi_{x,t}, \eta_{x,t})
e^{\textstyle\sum_x(\psi^*_{x,t} \psi_{x,t} + \eta^*_{x,t}   \eta_{x,t})}
\label{eq13}
\end{equation}
which is true of any normal ordered function $F$ of the operators
$a_x^\dag,b_x^\dag, a_x,b_x$. 

As we indicated above, the Hamiltonian is in normal order form
provided one separates the local term $e^2 V_{xx} q_x q_x$ 
in the Coulomb Hamiltonian into two normal-ordered pieces
\begin{equation}
e^2 V_{xx} q_x q_x = e^2 V_{xx} :q_x q_x: + \; 
e^2 V_{xx} (a^\dag_x a_x + b^\dag_x b_x).
\label{eqNormal}
\end{equation}
By reassigning the quadratic term in Eq.~\ref{eqNormal} to $H_2$, the
exponent $-H \,\delta$ in Eq.~\ref{eq12} is normal ordered.
The exponential $\exp(-H \,\delta)$ is not normal-ordered, but it differs
from its normal ordered form by terms $O(\delta^2)$, which give
a vanishing contribution to the integral in the limit of $N_t \to \infty$. 
Neglecting these terms we may replace the operator
expression $\exp(-H \,\delta)$ with an exponential involving the
fermionic fields, as follows from Eq.~\ref{eq13}. 
We thus obtain the following expression for the partition function 
\begin{equation}
Z  = \lim_{N_t\to \infty} \int \prod_{m} d\psi_m^* d\psi_m d\eta_m^* d\eta_m 
\times e^{\textstyle-\sum_{m,n}( \psi^*_m M_{m,n} \psi_n+\eta^*_m M_{m,n} \eta_n)} 
e^{\textstyle-\sum_{x,y,t} e^2 Q_{x,t} V_{x,y} Q_{y,t} \delta } \label{eq14}
\end{equation}
where $Q_{x,t} =\psi^*_{x,t} \psi_{x,t} - \eta^*_{x,t} \eta_{x,t}$ and
we have used $m$ (and $n$) as a shorthand for the indices $x,t$.
$M$ is a matrix whose components may be deduced from
\begin{equation}
\sum_{m,n}  \psi^*_m M_{m,n} \psi_n  = 
\sum_t\Big[\sum_x \psi^*_{x,t} (\psi_{x,t+1}-\psi_{x,t})
+ e^2 V_{xx} \psi^*_{x,t} \psi_{x,t} \, \delta  - 
\kappa \sum_{\langle x,y\rangle} 
(\psi^*_{x,t}\psi_{y,t}+\psi^*_{y,t}\psi_{x,t})\,\delta\Big]
\label{eq15}
\end{equation}
where $\psi_{x,N_t}$ must be identified with $-\psi_{x,0}$.

We now perform  a Hubbard-Stratonovich transformation, introducing
c-number real variables $\phi_{x,t}$ to recast the exponential
with the quartic term in the form
\begin{equation}
e^{\textstyle-\sum_{x,y,t} e^2 Q_{x,t} V_{x,y} Q_{y,t} \delta } = \int \prod_{x,t} d\phi_{x,t} 
e^{\textstyle-\sum_{x,y,t} \phi_{x,t} (V^{-1})_{x,y} \phi_{y,t} 
\delta/4}e^{\textstyle-\sum_{x,t} \imath e \phi_{x,t}Q_{x,t}\delta },
\label{eq16}
\end{equation}
where we have absorbed a constant measure factor in the definition
of the integral over $\phi_{x,t}$.  
Inserting the r.h.s.~of Eq.~\ref{eq16} into Eq.~\ref{eq14} 
and introducing the diagonal matrix $\Phi_{x, t; y, \tau} = 
\left(\phi_{x, t}\delta\right) \delta_{x,y}\delta_{t,\tau}$
produces the compact result
\begin{equation}
Z = \int  d\phi d\psi^* d\psi d\eta^* d\eta 
e^{\textstyle- \phi V^{-1} \phi  \delta/4 - \psi^* (M+\imath e \Phi) \psi 
-\eta^* (M-\imath e \Phi) \eta}
\label{eq19}
\end{equation}
where we have used matrix notation for all the sums and have dropped
the limit notation.

The Gaussian integration over the anticommuting variables can now
be done to obtain
\begin{equation}
Z=  \int  d\phi e^{\textstyle- \phi V^{-1} \phi  \delta/4}
{\rm det}(M-\imath e \Phi)
{\rm det}(M+\imath e \Phi).
\label{eq20}
\end{equation}
Because of the identity,
\begin{eqnarray*}
\det(M-\imath e \Phi)\det(M+\imath e \Phi) = 
\det[(M+\imath e \Phi)^\dag (M+\imath e \Phi)]
\end{eqnarray*}
the measure is positive definite. The down spins are treated
as antiparticles (holes) moving backward in time relative to 
the up spins, exactly canceling the phase for each separately.
Correlators for the fermion operators are now obtained by
integrating the appropriate matrix elements of $(M\pm\imath e \Phi)^{-1}$
with the measure given by Eq.~\ref{eq20}.

Equation~\ref{eq20} is the main result of our work.  It establishes
the partition function and expectation values as integrals over
real variables with a positive definite measure.  This is a crucial step
for the application of stochastic approximation methods.  There
remains the problem of sampling the field $\phi_{x,t}$ with a measure
which contains the determinant of a large matrix.  But, following
what is done in lattice gauge theory, this challenge
can be overcome through the application of the hybrid
Monte Carlo (HMC) technique~\cite{HMC}.  In a broad outline, in HMC
one first replaces the determinants in Eq.~\ref{eq20} with a Gaussian 
integral over complex pseudofermionic variables $\zeta_{x,t}$:
\begin{equation}
\det  \big[(M+\imath e \Phi)^\dag (M+\imath e \Phi)\big]
= \int  d\zeta^* d\zeta e^{\textstyle- \zeta^* (M+\imath e \Phi)^{\dag \, -1 } 
(M+\imath e \Phi)^{-1} \zeta}.
\label{eq21}
\end{equation}
(In this equation and in the following Eq.~\ref{eq22} we absorb 
an irrelevant, constant measure factor in the integrals.)
One then introduces real ``momentum variables'' $\pi_{x,t}$
conjugate to $\phi_{x,t}$ and inserts in Eq.~\ref{eq21} unity written
as a Gaussian integral over $\pi$.  One finally arrives at
\begin{equation}
Z =  \int  d\phi d\pi d\zeta^* d\zeta 
e^{\textstyle- \phi V^{-1} \phi  \delta/4 - \zeta^* (M+\imath e \Phi)^{\dag \, -1} 
(M+\imath e \Phi)^{-1} \zeta - \pi^2/2}.
\label{eq22}
\end{equation}
The idea of HMC  is to consider the simultaneous distribution
of the variables $\phi, \pi, \zeta$ and $\zeta^*$ determined by the
measure in Eq.~\ref{eq22}.  The phase space of these variables
is explored by first extracting the $\pi$, $\zeta$ and $\zeta^*$ 
according to their Gaussian measure, and then evolving the $\phi$
and $\pi$ variables with fixed $\zeta, \zeta^*$ according to 
the evolution determined by the Hamiltonian
\begin{eqnarray*}
{\cal H}(\pi,\phi)=\frac{\pi^2}{2}   + \frac{\phi V^{-1} \phi  \delta}{4} +
\zeta^* (M+\imath e \Phi)^{ \dag \, -1 } (M+\imath e \Phi)^{-1} \zeta.
\end{eqnarray*}
Because of Liouville's theorem, the combined motion through phase
space produces an ensemble of variables distributed according
to the measure in Eq.~\ref{eq22} and, in particular, of fields $\phi$ 
distributed according to Eq.~\ref{eq20}.

Of course, the discussion above assumes that the Hamiltonian evolution of
$\phi$ and $\pi$ is exact, which will not be the case with a numerical
evolution.  The HMC algorithm addresses this shortcoming
 1) by  approximating the evolution with a symplectic integrator which
is reversible and preserves phase space, 2) by performing a Metropolis 
accept-reject step at the end of the evolution, 
based on the variation of the value of the Hamiltonian.

\section{Numerical tests}

We tested our method on the two-site system
obtained by taking $L=1$, which can be solved exactly.  We label
the sites $x=0, 1$. With $\kappa =1/3$, the Hamiltonian
$H = H_2 + H_C$ is now
\begin{eqnarray}
H_2 & = & -(a_1^\dag a_0 + a_0^\dag a_1 + b_1^\dag b_0 + b_0^\dag b_1) 
+ \mu(a^\dag_x a_x + b^\dag_x b_x) \nonumber \\
H_C & = & 2 e^2(a_0^\dag a_0 - b_0^\dag b_0)(a_1^\dag a_1 - b_1^\dag b_1) 
+ \frac{2  e^2}{r_0} a_x^\dag b_x^\dag a_x  b_x
\end{eqnarray}
where we have taken $V_{0,1}=V_{1,0}=1/3$ and a local interaction term
$V_{0,0}=V_{1,1}=1/r_0$.  The radius $r_0$ sets the physical scale
in lattice units for localization of the net charge at the carbon atom.
It must be restricted to $r_0 < 1$ for stability of the vacuum. Also the normal
ordering prescription for $e^2 V_{xx}q_x q_x$ in Eq.~\ref{eqNormal}
adds a new contribution to $H_2$ in the form of an $J_3$ ``chemical
potential'' $\mu a^\dag_{x,s} \sigma^{ss'}_3 a_{x,s'}$.  It is well
known~\cite{iso} that a $J_3$ chemical potential for any value of
$\mu$ does not introduce a phase in the measure.  To maintain the
full $SU(2)$ ``flavor'' symmetry of the tight-binding graphene
Hamiltonian, we must set $\mu = e^2/r_0$ to its proper value.  For the two-site
system, the spin generators of Eq.~\ref{eq:iso} become
\begin{eqnarray*}
J_+ = J^\dag_- = (-1)^x a^\dag_x b^\dag_x 
\mbox{ and } J_3 = [a^\dag_x a_x + b^\dag_x b_x]/2 -1,
\end{eqnarray*}
allowing us to unambiguously classify the 16 states as
5 singlets, 4 doublets and one triplet, given in Table~\ref{tab:multiplets}.

\begin{table}[h]
\centering
\begin{tabular}{|l|r|r|}
\hline
J & eigenstates $| n \rangle$ & Eigenvalues E   \\
\hline
J = 1  &  $\vert 0 \rangle \quad , \quad 
  \frac{a_0^\dag b_0^\dag -a_1^\dag b_1^\dag}{\sqrt{2}}\vert 0 \rangle \quad , \quad 
a_0^\dag a_1^\dag b_0^\dag b_1^\dag\vert 0 \rangle$
& $0$  \\
\hline
J = 1/2         &$\frac{a_0^\dag-a_1^\dag}{\sqrt{2}}\vert 0 \rangle \quad , \quad 
\frac{a_0^\dag-a_1^\dag}{\sqrt{2}}b_0^\dag b_1^\dag \vert 0 \rangle $
& $3\kappa+ e^2/r_0$  \\
     &$\frac{b_0^\dag-b_1^\dag}{\sqrt{2}} \vert 0 \rangle\quad , \quad 
\frac{b_0^\dag-b_1^\dag}{\sqrt{2}} a_0^\dag a_1^\dag \vert 0 \rangle$
& $3\kappa+ e^2/r_0$ \\
        & $ \frac{a_0^\dag+a_1^\dag}{\sqrt{2}}\vert 0 \rangle \quad,\quad
\frac{a_0^\dag+a_1^\dag}{\sqrt{2}}b_0^\dag b_1^\dag\vert 0 \rangle $
& $-3\kappa+e^2/r_0$  \\
        &$\frac{b_0^\dag+b_1^\dag}{\sqrt{2}} \vert 0 \rangle\quad , \quad 
\frac{b_0^\dag+b_1^\dag}{\sqrt{2}} a_0^\dag a_1^\dag \vert 0 \rangle$
& $-3\kappa+ e^2/r_0$  \\
        &$ a_0^\dag a_1^\dag \vert 0 \rangle \qquad , \qquad 
b_0^\dag b_1^\dag \vert 0 \rangle$
& $2 e^2 + 2 e^2/r_0$  \\
\hline
J = 0         &$\frac{(a_0^\dag b_0^\dag +a_1^\dag b_1^\dag)\cos\theta
-(a_0^\dag b_1^\dag +a_1^\dag b_0^\dag)\sin\theta }{\sqrt{2}}\vert 0 \rangle$
& $\sqrt{36\kappa^2+(1- 1/r_0)^2e^4}  -  e^2 +  e^2/r_0$ \\
        &$\frac{(a_0^\dag b_0^\dag +a_1^\dag b_1^\dag)\sin\theta
+(a_0^\dag b_1^\dag +a_1^\dag b_0^\dag)\cos\theta }{\sqrt{2}}\vert 0 \rangle$
& $-\sqrt{36\kappa^2+(1-1/r_0)^2e^4} -  e^2 +  e^2/r_0$  \\
        &$\frac{a_0^\dag b_1^\dag -a_1^\dag b_0^\dag}{\sqrt{2}}\vert 0 \rangle$
& $- 2 e^2 + 2 e^2/r_0$   \\
\hline
\end{tabular}
\caption{Spin multiplets and energies for the 2 site Graphene model.}
\label{tab:multiplets}
\end{table}

\begin{figure}[h!]
\centering
\includegraphics[width=0.6\textwidth]{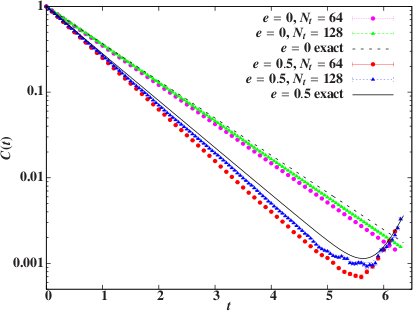}
\caption{HMC results for $C_a(t)$ (Eq.~5.2) with $e = 0$
and $e = 0.5$ compared to the exact correlators.
Here $\mu = e^2/r_0$, $r_0 = 1/2$ and $\beta = N_t \delta = 6.4$.}
\label{fig:figure5}
\end{figure}

We compared HMC results for expectation values of several
products of fermionic operators with the corresponding
exact values, finding satisfactory agreement.
For example, the correlation function
\begin{equation}
C_a(t)=\langle (a_0-a_1)(t) \, (a_0^{\dag}-a_1^{\dag})(0)\rangle /2
\label{eq25a}
\end{equation}
is illustrated in Figure.~\ref{fig:figure5}, which shows HMC results converging
to the exact correlators for both the free theory with $e = 0$
and an interacting case with $e = 0.5$.

\begin{figure}[t]
\begin{minipage}[b]{0.47\linewidth}
\centering
\includegraphics[width=7cm]{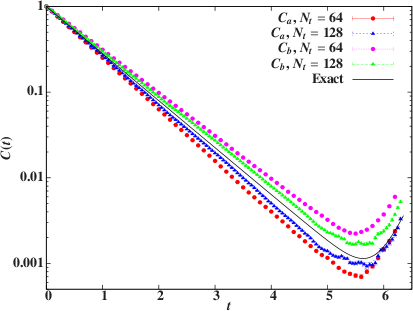}
\caption{HMC results for $C_a(t)$ (Eq.~5.2) and $C_b(t)$ (Eq.~\protect\ref{eq25b}) 
with $e = 0.5$ compared to the exact correlator.  As in  Figure.~\protect\ref{fig:figure5}, 
$\mu = e^2/r_0$, $r_0 = 1/2$ and $\beta = N_t \delta = 6.4$.}
\label{fig:figure6}
\end{minipage}
%\hspace{0.5cm}
\hspace{0.01\linewidth}
\begin{minipage}[b]{0.47\linewidth}
\centering
\includegraphics[width=7cm]{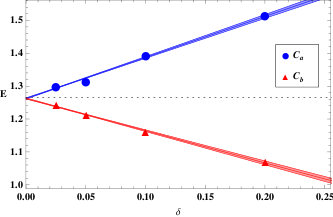}
\caption{Linear extrapolation (with error band) of the HMC energies for
doublet correlators $C_a$ and $C_b$ as a function of the ``time'' 
lattice spacing $\delta = 6.4/N_t$ for $N_t = 32$, $64$, $128$ and $256$.
The dotted horizontal line marks $E_0 = 1.266$.}
\label{fig:figure7}
\end{minipage}
\end{figure}

A stringent test is to demonstrate the convergence to exact $SU(2)$
symmetry in the ``time'' continuum limit.  To this end, consider a
second correlation function,
\begin{equation}
C_b(t)=\langle (b_0^{\dag} + b_1^{\dag})(t) \, (b_0 + b_1)(0)\rangle /2 \; ,
\label{eq25b}
\end{equation}
related to $C_a(t)$ by an $SU(2)$ rotation. Figure.~\ref{fig:figure6}
illustrates that HMC results for both $C_a(t)$ and $C_b(t)$
converge to the same continuum limit.

We extract the energies of the doublet states at nonzero
$\delta = \beta/N_t$ by fitting the correlator data in
Figure~\ref{fig:figure6} to single exponentials, $C(t) \approx e^{\textstyle-Et}$ for
fit range $0.4 < t < 4$. The results in Figure.~\ref{fig:figure7} clearly show
linear behavior $E \approx E_0 + c_1\delta$, converging to the exact continuum
$E_0 = e^2 + \sqrt{4 + e^4} - 1 \approx 1.266$.
The continuum limit is consistent with restoration of the $SU(2)$
symmetry of the Hamiltonian: a joint linear fit to both sets of energies
gives $\lim_{\delta \to 0} E = 1.262 \pm 0.004$, with $c_1 = 1.25 \pm 0.07$
and $-0.98 \pm 0.04$ for correlators $C_a$ and $C_b$, respectively.

\section{Discussion}

Before scaling up our simulations to large lattices, it is useful to
reflect on the formalism presented here. In our formulation, the
theory is reduced to a 2+1 dimensional lattice theory with no sign
problem. The non-local potential when expressed as a convolution can be
computed efficiently using a Fast Fourier transform.  The
only approximation to the tight binding Hamiltonian for the nearest
neighbor coupling is the discretization of the Euclidean time, $\delta$. So as
we extrapolate to $\delta \rightarrow 0$, we can in principle solve
the exact tight binding model numerically to arbitrary precession.
However there are, as in all lattice field theories, improved
discretization and algorithms that may accelerate the convergence. The
difference here is the ``continuum limit'' involves only the ``time''
discretization and there is no symmetry rotating this axis into the
hexagonal plane. So we are exploring alternative discretization in ``time'' which may
improve  computational efficiency and accuracy for small  $\delta$.

On a theoretical level, it is instructive to re-derive the
discretization for finite lattice spacing $\delta$ starting with our
exact time continuum ($\delta = 0$) acton or its Lagrangian ($S= \int dt {\cal L}$)
derived above,
\begin{equation}
{\cal L}(t) = \psi^\dag_x(t) (\partial_t + i e \sigma_3 \phi_x(t) ) \psi_x(t) + e^2 V_{xx} \psi^\dag_x(t) \psi_x(t) -  \kappa \sum_{ \langle x, y\rangle} \psi^\dag_x(t) \psi_y(t) 
+ \frac{1}{ 4} \phi_{x}(t)  V^{-1}_{xy} \phi_{y}(t)  \; ,
\label{eq:Lagrangian}
\end{equation}
where for simplicity we have combined the two spin components
$\psi,\eta$ into a spinor field: $\psi^\dag_x = (\psi^*_x,
\eta^*_x)$ and $\psi_x = (\psi_x, \eta_x)^T$. Let us also introduce
a a staggered mass parameter, $+/-m \psi^\dag_x \psi_x $ for sites on
the A/B lattices respectively, which will be useful in numerical
simulation in any case due to critical slowing down in the chiral
limit. The result for the free theory ($e = 0$) on the nearest
neighbor hexagonal lattice is to lift the zero mode symmetrically to
give the continuum dispersion relation,
\be
E = \pm i \sqrt{ \omega^2_k + m^2} \; .
\label{eq:DR}
\ee
At $m = 0$ this is just the well know spectrum of the nearest neighbor graphene
Hamiltonian $H_2$ by going to Fourier space. Using the
oblique reciprocal co-ordinates, $k_1 = 3 k_x/2+ \sqrt{3} k_y/2, k_2 = \sqrt{3}
k_y$, on the Hexagonal lattice described in the caption of
Figure.\ref{fig:figure6}, $\omega_k$ is given by $\omega_k = | 1 +
e^{\textstyle i k_1} + e^{\textstyle i k_2}|$ in spatial units set by
taking $\kappa = 1$. On a finite periodic lattice, the momenta
are discrete ($k_i = 2 \pi n_i/L_i$ for integer $n_i$), so to preserve
the exact zero modes at the apex of the Dirac cones, $k_1 = - k_2 = \pm 2 \pi/3$, we must chose
$L_i$ to be divisible by 3.

Now let us  re-introducing the temporal lattice spacing,
$\delta$, into our continuum Lagrangian (\ref{eq:Lagrangian}). Our
derivation above amounted to replacing the derivative by a forward
difference, $\partial_t \psi_x \rightarrow (\psi_{x,t+1} -
\psi_{x,t})/\delta$, but there are other equally valid discretizations
with the same continuum limit. Another attractive choice is to use this
forward difference on the A lattice, $\partial_t \psi_x \rightarrow
(\psi_{x,t+1} - \psi_{x,t})/\delta$, but a backward difference on the B
lattice, $\partial_t \psi_x \rightarrow (\psi_{x,t} -
\psi_{x,t-1})/\delta$, which has an interesting lattice symmetry of
time reversal times A/B lattice exchange. Comparing the free ($e = 0$) dispersion
relation at finite $\delta$, we have 
\begin{equation}
  (2/\delta) e^{\textstyle i E \delta/2} \sin(E \delta /2) = \pm i  \sqrt{ \omega^2_k + m^2} \; 
\end{equation}
for our original choice of all forward differences versus 
\begin{equation}
(2/\delta) \sin(E \delta /2) = \pm i \sqrt{\frac{\omega^2_k + m^2}{1 - m \delta}} \; . 
\end{equation}
for the A/B alternating forward/ backward difference choice. Both dispersion relations of
course yield the well know exact continuum dispersion relation
(\ref{eq:DR}) for $\delta \rightarrow 0$. However, the second one is
accurate to $O(\delta^2)$ and has a symmetric spectrum in the complex
plane characteristic of a relativistic fermion in the continuum. Very
likely this form has some real advantages which we are actively
investigating.

Next let us return to the interacting case. Note in our Lagrangian
(\ref{eq:Lagrangian}), that the quadratic (kinetic) action for the
electrons exhibits a 1-d space independent ``gauge invariance''
\begin{equation}
 \psi_x(t) \rightarrow \exp[- i e \sigma_3 \theta(t)] \psi_x(t) \; , \quad \psi^\dag_x(t)  \rightarrow \psi^\dag_x(t) \exp [ i e \sigma_3 \theta(t)]  \; , \quad \phi_x(t) \rightarrow \phi_x(t) + \partial_t \theta(t) \; ,
\end{equation}
much like the residual gauge invariance of the 4-d representation with
gauge potential $A_0$. However $\phi$ is not the same object. The
potential term in ${\cal L}$ is not gauge invariant and thus it fixes
this residual gauge transformation. It is natural to ask what happens
if we preserve this gauge invariance for the kinetic term on the
temporal lattice. We find an intriguing connection between the normal
ordering term, $e^2 V_{xx} \psi^\dag_x \psi_x$, and the gauge
invariance of the kinetic term, which suggest another discretization
using compact gauge variables on the links,
\be
\int dt  \psi^\dag_x(t) (\partial_t + i e \phi_x(t) \sigma_3) \psi_x(t) \rightarrow \sum_t \psi^\dag_{x,t} (e^{\textstyle i e \delta \sigma_3 \phi_{x,t}} \psi_{x,t+1}  - \psi_{x,t}) \; ,
\ee
which like Wilson link variables in lattice field theory preserve
gauge invariance of the kinetic term at the discrete times $t$  for each
temporal slice. Now let us expand in $\delta$ as usual to see how
this approaches the continuum Lagrangian (\ref{eq:Lagrangian} ): 
\be
\sum_t \psi^\dag_{x,t} (e^{\textstyle i e \delta \sigma_3 \phi_{x,t}} \psi_{x,t+1}  - \psi_{x,t}) \simeq \int dt \psi^\dag \partial_t \psi_x + i e \int dt \psi^\dag_x \sigma_3
\phi_{x} \psi_{x} - \delta \int dt\frac{e^2}{2} \phi^2_{x} \psi^\dag_{x} \psi_{x} + \cdots 
\label{eq:segull}
\ee
The first and two second term on the right hand side are the required form to account for the
Hubbard-Stratonovich transformation. Usually we would drop the third
term as irrelevant in the continuum. However the Hubbard-Stratonovich
field, $\phi_{x,t}$, is not a dynamical field and on the lattice its
fluctuations are controlled only by the quadratic potential in the
action,
\be
\frac{1}{ 4}\int dt  \phi_{x}(t)  V^{-1}_{xy} \phi_{y}(t) \rightarrow  
\sum_t  \frac{\delta }{4} \phi_{x,t} V^{-1}_{xy} \phi_{y,t}    \; ,
\ee
This term implies that the variance for small $\delta$  diverges on each time slice:
$\langle \phi^2 \rangle= O(1/\delta)$. Consequently the third term is
not suppressed and must be taken into account. Remarkably it can be
shown that this term is exactly equivalent to adding the normal
ordering term to leading order in $\delta$. To show this we make a
field redefinition: $\phi_x \simeq \widetilde \phi_x - \delta V_{xy}
\widetilde \phi_y \psi^\dag_y \psi_y$ which when substituted into the
lattice Lagrangian cancels the $\phi^2_x$ term in Eq.~\ref{eq:segull}  to leading order but
 to
preserve the measure in the path integral also requires us to  introduce the
Jacobian of this change of variables. This Jacobian gives rise
to our normal ordering term in the continuum limit,
\be
\int {\cal D} \phi_{x,t} = \int {\cal D} \widetilde \phi_{x,t} \; \det\big[\frac{\partial \phi_{x,t}}{\partial \widetilde \phi_{x',t'}} \big] = 
\int  {\cal D} \widetilde \phi_{x,t} \;  e^{ \textstyle Tr \log(1 - \delta V \psi \psi)} \simeq \int  {\cal D} \widetilde\phi_{x,t} e^{ \textstyle -   \int dt  V_{xx} \psi^\dag_{x}(t) \psi_x(t)} \; .
\ee
In summary an alternative to the normal ordering term on the lattice
is to ignore it and introduce compact gauge links. 

An independent derivation  that yields  the same result  is to reverse the argument in Sec.~\ref{sec:pathIntegral} by first introducing the  Hubbard-Stratonovich transformation in the Hamiltonian  for each of the $N_t = \beta/\delta$factors in the discretized transfer matrix,
\begin{equation}
e^{\textstyle-\beta H} =  e^{\textstyle- H\, \delta} e^{\textstyle - H\, \delta} \dots e^{\textstyle- H\, \delta} \quad (N_t\;  \mbox{terms}) \; .
\label{eq11}
\end{equation}
Since the local charges commute $[q_x, q_y] =0$ the result of the Hubbard-Stratonovich transformation   to leading order is simply the 
replacement, 
\begin{equation}
e^{ \textstyle -  H\,  \delta} =  \int \Pi_x d\phi_{x,t} e^{ \textstyle - H_2 \, \delta- i e \delta \phi_{x,t} q_x + \frac{\delta }{4} \phi_{x,t} V^{-1}_{xy} \phi_{y,t}} \; ,
\end{equation}
for each factor on time slice $t$. Now if we use the exact normal ordering identity, 
\be \exp[
- i e \delta \phi_x a^\dag_x \sigma_3 a_x] \rightarrow \exp[\psi^\dag_x e^{ \textstyle - i e \delta \sigma_3 \phi_x } \psi_x ]
\; , 
\ee
given as Eq. A8 in Ref.~\cite{Luscher:1976ms} and expand to leading
order, we again get the quadratic (seagull) term, $ - (e^2/2) \delta^2 \phi^2_x \psi^\dag_x \psi_x$, in Eq.~\ref{eq:segull} . To formally take the continuum limit with this
term of course we  must again make a change of variables arriving
to the unique continuum  Lagrangian (\ref{eq:Lagrangian}). 
This alternate derivation makes it clear that the two lattice expressions (with 
$ - (e^2/2) \delta^2 \phi^2_x \psi^\dag_x \psi_x$ or $e^2 V_{xx} \psi_x \psi_x$)  are simply different
approximation to account for  normal ordering. What is surprising is that compact
link variables also provide this contribution and that in this form
there is no explicit dependence on $V_{xx}$ need to define the effects of normal ordering.  The dependence is implicit in
the necessity to  stabilize the quadratic form with a local repulsion at the
Carbon ions to allow the Hubbard-Stratonovich transformation to be legitimate. 

Beyond the curiosity of this seagull term inducing the
normal ordering correction, it opens up the attractive option of using
compact links, as if $\phi$ were a gauge potential. One might suppose
that this is a consequence of dimensional reduction from a real 4-d
gauge theory but it is also more general since the result is exact for
any static potential term $V_{xy}$.  The generality of the
construction allows any static potential and any fixed geometry
for the graphene sheet. It is also possible to include fluctuation of the graphene
sheet by modeling phonon interactions with additional dynamical spatial gauge links. 
At present
we are investigating a range of discretization schemes numerically in
comparison with  solving the  exact spectrum on lattices with $L_1 = L_2 = 2$ to see
how each converges to the continuum limit. These test lattices have 8
sites with two spins per site so in total $2^{16} =65536$ states offer a rigorous and  non-trivial
verification of our HMC algorithm and software.

\section{Conclusion}
We have presented a Lagrangian  formalism for the nearest neighbor 
tight binding theory for single layer graphene that allows one to 
do Hybrid Monte Carlos simulations by the traditional method
of Euclidean lattice field theory.  

While the results we reported are for a small single hexagonal cell test systems, they demonstrate
that HMC simulations of graphene directly
on the hexagonal graphene lattice are possible and have the potential to
produce valuable results.  The dominant nearest neighbor hopping term
has no sign problem, and we anticipate that
a small next-to-nearest neighbor coupling $\kappa'/\kappa \simeq 0.03$
can be accommodated by reweighting without a prohibitive cost. The crucial
observation is the cancellation between the phase of the up spin and
down spin determinant, when the latter are treated as holes moving
backward in time.  The only approximation is the
discretization error introduce by a lattices spacing $\delta$ for the ``time'' or temperature lattice.
This opens up a method for effectively solving this tight binding model numerically
to arbitrary precision subject to taking the 1-d continuum limit. Relative
to computational cost of lattice QCD simulations, clearly lattice graphene  simulations are   not only
 feasible but considerably more tractable  with modern computing 
resources and algorithms. Still this is a young computational field and there is much work
to be done to explore both algorithms and discretization 
scheme to optimize these simulations. 

  We are currently pursuing simulations
of larger systems, and beginning to explore the many possible
generalizations such as  distortions of the lattice, phonons, inclusion of magnetic fields, etc.
We have a fully functioning  code for our periodic hexagonal lattice 
with nearest neighbor hopping terms and a periodic Coulomb potential.
We are putting the conjugate gradient inverter and the molecular dynamics
evolution onto single GPUs. Since the NVIDIA Tesla class GPU have  device
memories in the range of 3 to 6 GigaBytes, this will allow
a very substantial on card lattice volumes  and a  performance of O(100)  Gigaflops per GPU.
The potential problems this open up to simulation are  substantial. Graphene
has a range of interesting properties. Our first goal  is to repeat the
calculation in Ref.~\cite{LGT-graphene, Lahde:2011yu}   of the critical charge for the excitation gap (or chiral symmetry breaking) 
for single layer graphene due to the strong Coulomb potential. Having a
reliable number  for this critical parameter for  the tight binding model is of considerable experimental
interest.  Beyond that there are a large range of effects due to changing boundary conditions
and the introduction of phonons, which are easily introduced into the
lattice Lagrangian formalism presented here.  It is even feasible to 
simulate small graphene samples with lattices, identical in size and geometry to those being
studied in experimental investigation. Here finite size effects are real and very interesting,
both theoretically and technologically.

\paragraph{\textbf{Acknowledgments:}} We wish to acknowledge the many fruitful
conversations with Dr.~Ronald Babich, Prof.~Antonio Castro Neto,
Prof.~Claudio Chamon and Dr.~MIchael Cheng during the course of this research and support
under DOE grants DE-FG02-91ER40676,
DE-FC02-06ER41440, and NSF grants OCI-0749317, OCI-0749202. Part of
this work was completed while two of the authors were at the Aspen
Center for Physics.

%\newpage


\begin{thebibliography}{99}
\bibitem{graphene}   K.~S.~Novoselov {et al.},
  ``Electric Field Effect in Atomically Thin Carbon Films'',
  {\em Science} {\bf 306}, 666 (2004).

\bibitem{dispersion}
  A.~H.~Castro Neto {et al.},
  ``The electronic properties of graphene'',
  {\em Rev. Mod. Phys.} {\bf 81}, 109 (2009).% [arXiv:0709.1163 [cond-mat.other]].

\bibitem{LGT}
  H.~J.~Rothe,
  {\em Lattice Gauge Theories: An Introduction},
  (World Scientific, Singapore, 2005).%, 605 pages (World Scientific Lecture Notes in Physics, 74).

\bibitem{LGT-graphene}
  J.~E.~Drut and T.~A.~L\"ahde,
  ``Is Graphene in Vacuum an Insulator?'',
  {\em Phys. Rev. Lett.} {\bf 102}, 026802 (2009); % [arXiv:0807.0834 [cond-mat.str-el]];
  % ``Lattice field theory simulations of graphene'',
  {\em Phys. Rev. B} {\bf 79}, 165425 (2009).% [arXiv:0901.0584 [cond-mat.str-el]].

%\cite{Lahde:2011yu}
\bibitem{Lahde:2011yu}
  T.~A.~Lahde and J.~E.~Drut,
  ``Strongly Coupled Graphene on the Lattice,''
  arXiv:1111.0929 [hep-lat].
  %%CITATION = ARXIV:1111.0929;%%


\bibitem{HMC}
  Simon Duane {et al.},
 ``Hybrid Monte Carlo''
  {\em Phys. Lett. B} {\bf 195}, 216 (1987).


%\cite{Giedt:2009td}
\bibitem{Giedt:2009td}
  J.~Giedt, A.~Skinner and S.~Nayak,
  ``Effects of flavor-symmetry violation from staggered fermion lattice
  simulations of graphene,''
  arXiv:0911.4316 [cond-mat.str-el].
  %%CITATION = ARXIV:0911.4316;%%

\bibitem{Wehling:2011}
T.O. Wehling, E. Sastioglu, C> Friedrich, A> I. Lichtenstein,  M. I. Katsnelson andS. Blugel,
``Strength of Effective Coulomb Interactions in Graphene and Graphite'' 
{\em Phys.  Rev.  Lett.} {\bf 106}, 236805 (2011).

\bibitem{Negele-Orland}
  J.~W.~Negele and H.~Orland,
  {\em Quantum Many Particle Systems},
  (Addison-Wesley, Redwood City, California, 1988).%, 459 pages (Frontiers in Physics, 68).

\bibitem{HST}
  R.~L.~Stratonovich,
  ``On a Method of Calculating Quantum Distribution Functions'',
  {\em Sov. Phys. Dokl.} {\bf 2}, 416 (1958);
  J.~Hubbard,
 ``Calculation of Partition Functions'',
  {\em Phys. Rev. Lett.} {\bf 3}, 77 (1959).

\bibitem{BSS}
R. Blankenbecler, D. J. Scalapino and R. L. Sugar,
``Monte Carlo calculations of coupled boson-fermion systems. I''
{\em Phys. Rev. D} {\bf 24}, 2278 (1981).

\bibitem{iso}
  M.~G.~Alford, A.~Kapustin and F.~Wilczek,
  ``Imaginary chemical potential and finite fermion density on the lattice'',
  {\em Phys. Rev. D} {\bf 59}, 054502 (1999); % [arXiv:hep-lat/9807039]. 
  D.~T.~Son and M.~A.~Stephanov,
   ``QCD at finite isospin density'',
  {\em Phys. Rev. Lett.} {\bf 86}, 592 (2001).% [arXiv:hep-ph/0005225].


%\cite{Luscher:1976ms}
\bibitem{Luscher:1976ms}
  M.~Luscher,
  ``Construction Of A Selfadjoint, Strictly Positive Transfer Matrix For
  Euclidean Lattice Gauge Theories,''
  Commun.\ Math.\ Phys.\  {\bf 54}, 283 (1977).
  %%CITATION = CMPHA,54,283;%%


\end{thebibliography}
\end{document}